\documentclass[aps,pra,reprint,showpacs,superscriptaddress,longbibliography]{revtex4-1}
\usepackage{mathtools,amssymb,graphicx,units}
\usepackage[plainpages=false,pdfpagelabels,colorlinks=true,linkcolor=red,urlcolor=blue,citecolor=blue,pdftitle={Title},pdfauthor={},pdfdisplaydoctitle=true,pdfduplex=DuplexFlipLongEdge]{hyperref}
\usepackage{soul} 
\usepackage[dvipsnames,usenames]{color}
\usepackage[normalem]{ulem}
\usepackage{graphicx}
\usepackage{color}
\usepackage{bbold} 
\usepackage{float}
\emergencystretch=\maxdimen
\usepackage{titlesec}

\newcommand{\iv}{\mathbf{i}}

\newcommand{\kv}{\mathbf{k}}

\begin{document}

\title{Superconducting Kondo phase in an orbitally-separated bilayer}
\author{Sebasti\~ao dos Anjos \surname{Sousa-J\'unior}} 
\email{sebastiaojr@pos.if.ufrj.br}
\affiliation{Instituto de F\'\i sica, Universidade Federal do Rio de Janeiro
Cx.P. 68.528, 21941-972 Rio de Janeiro RJ, Brazil}
\author{Jos\'e \surname{P. de Lima}}
\affiliation{Departamento de F\'\i sica, Universidade Federal do Piau\'\i , Teresina, PI, Brazil}
\author{Natanael C. Costa}
\affiliation{Instituto de F\'\i sica, Universidade Federal do Rio de Janeiro
Cx.P. 68.528, 21941-972 Rio de Janeiro RJ, Brazil}
\affiliation{International School for Advanced Studies (SISSA),
Via Bonomea 265, 34136 Trieste, Italy}
\author{Raimundo R. \surname{dos Santos}} 
\affiliation{Instituto de F\'\i sica, Universidade Federal do Rio de Janeiro
Cx.P. 68.528, 21941-972 Rio de Janeiro RJ, Brazil}

\begin{abstract}

The nature of superconductivity in heavy-fermion materials is a subject under intense debate, and controlling this many-body state is central for its eventual understanding. Here, we examine how proximity effects may change this phenomenon, by investigating the effects of an additional metallic layer on the top of a Kondo-lattice, and allowing for pairing in the former. 
We analyze a bilayer Kondo Lattice Model with an on-site Hubbard interaction, $-U$, on the additional layer, using a mean-field approach.
For $U=0$, we notice a drastic change in the density-of-states due to multiple-orbital singlet resonating combinations. 
It destroys the well-known Kondo insulator at half filling, leading to a metallic ground state, which, in turn, enhances antiferromagnetism through the polarization of the conduction electrons. 
For $U\neq 0$, a superconducting Kondo state sets in at zero temperature, with the occurrence of unconventional pairing amplitudes involving $f$-electrons. 
We establish that this remarkable feature is only possible due to the proximity effects of the additional layer. 
At finite temperatures we find that the critical superconducting temperature, $T_c$, decreases with the interlayer hybridization. 
We have also established that a zero temperature superconducting amplitude tracks $T_c$, which reminisces the BCS proportionality between the superconducting gap and $T_c$.

\end{abstract}


\maketitle


\section{Introduction}
\label{sec:intro}
The Kondo Lattice Model (KLM) \cite{Doniach1977,Lacroix1979} and its closely related Periodic Anderson Model (PAM) are believed to capture some of the basic aspects of magnetism in heavy-fermion materials~\cite{Tsunetsugu97,Coleman2007,Si2010,Fazekas1999}. 
Both models describe conduction electrons coupled to (quasi) localized $f$ moments, and the former may be viewed as the strong hybridization limit of the latter \cite{Schrieffer1966}. 
The screening of local moments by the conduction electrons favors a paramagnetic phase made up of singlets; competing with this, the polarization of the conduction electrons gives rise to a Ruderman-Kittel-Kasuya-Yosida (RKKY)  interaction \cite{Ruderman1954,Kasuya1956,Yosida1957}, which favors the formation of an antiferromagnetic state.
A quantum phase transition between these two tendencies takes place in the ground state, and many important physical concepts have emerged as a result of thorough investigations of these phase boundaries for the KLM, especially on a two-dimensional lattice~\cite{Coleman2007,Assaad1999,Watanabe2007,Asadzadeh13,Peters15,Li1996,Zhang2000,Zhang2010,Bernhard2015,Costa17a}. 

Another fascinating aspect of the heavy-fermion class of materials is the proximity of (and sometimes coexistence with) superconductivity and magnetic order in several compounds \cite{Pfleiderer2009}. 
From the theoretical point of view, a convenient starting point to model superconductivity  
in these materials is the KLM. 
Superconductivity in the standard KLM was found in the paramagnetic sector in the uncompensated regime through variational Monte Carlo simulations~\cite{Asadzadeh13}; it was also predicted by Dynamical Mean Field Theory (DMFT) calculations with special heavy fermion bands~\cite{Bodensiek13}. 
Alternative approaches based on added terms to the KLM, such as frustration due to second-neighbor hopping \cite{Asadzadeh14}, or by an extra orbital~\cite{Zegrodnik12} have also led to the formation of superconducting states; in both cases the presence of a direct Heisenberg interaction between local spins seems to provide the pairing `glue'. 

\begin{figure}[t]
\centering
\includegraphics[scale=0.36]{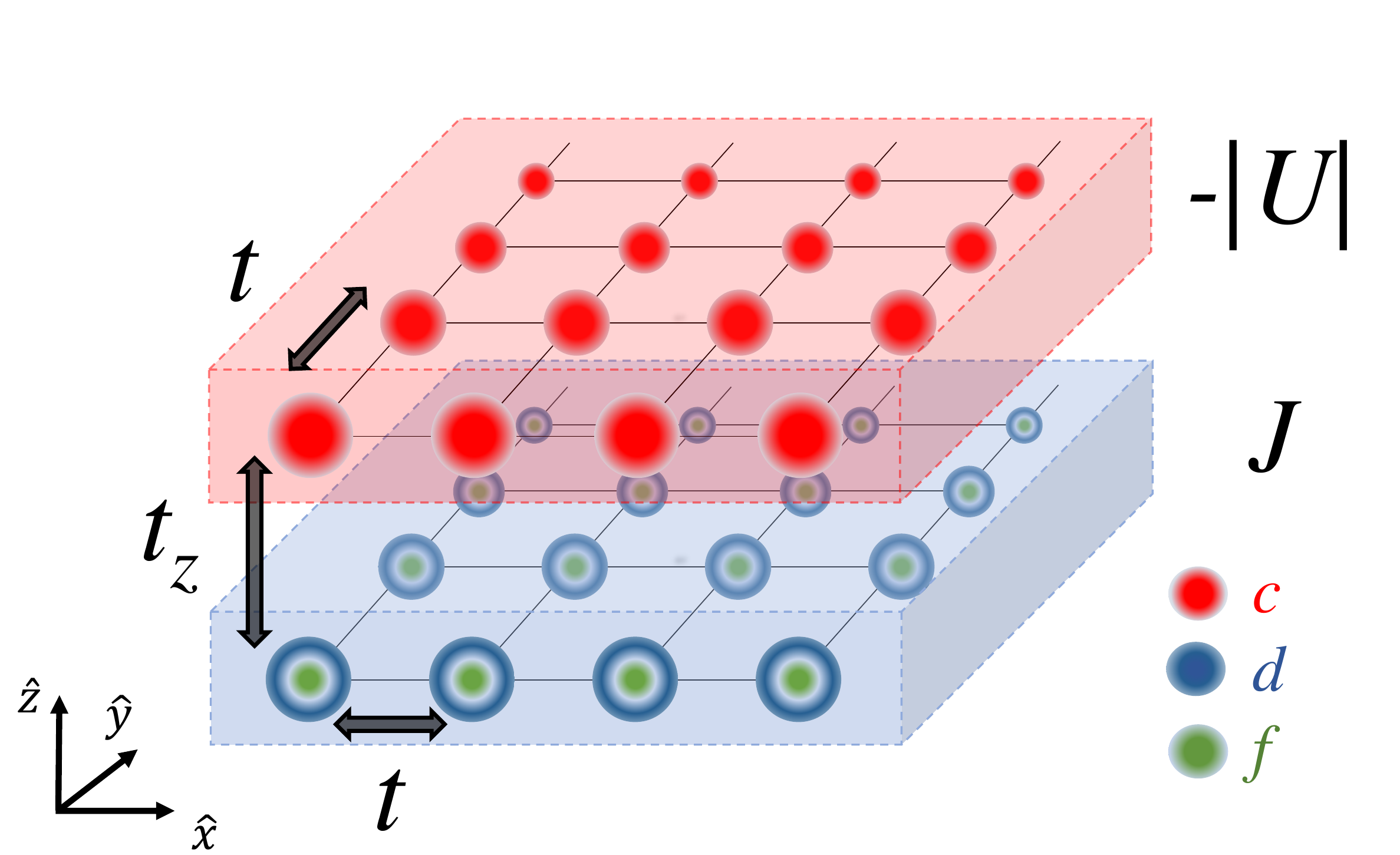} 
\caption{(Color online) 
Basic structure of our bilayer system: The $c$-electrons on the top layer may be subject to an on-site pairing interaction $-U<0$, while $d$ and $f$-electrons on the `Kondo-layer' are exchange-coupled (magnitude $J$).
The layers are hybridized through a hopping amplitude $t_{z}$, and all intra-layer hoppings are assumed equal to $t$.
}
\label{fig:bilayer}
\end{figure}

The quest for alternative scenarios leading to superconductivity in KLM-like models is therefore a question of current interest. 
In particular, we note that the Kondo singlet phase (both the insulator at half filling and the doped metallic phase) seems to be quite robust, since 
the inclusion of local pairing interactions between the conduction electrons only stabilizes superconductivity in the strong attractive coupling limit, or in the weak $d$-\!$f$ exchange coupling~\cite{Costa18a,Lechtenberg18}.  
One possible route towards superconductivity is the introduction of an extra layer, with the purpose of altering the band structure in a fundamental way.
Indeed, such strategy has proved fruitful, for instance in the case of Ce-based superlattices~\cite{Shishido10}, and more recently in the twisted bilayer graphene~\cite{Cao2018a,Cao2018b}, which revealed a wealth of interesting magnetic and superconducting phenomena.

With this in mind, here we investigate the effects of spatially separating Kondo and pairing physics by considering a bilayer in which the top layer favors superconductivity (through pairing of $c$-electrons) and the bottom layer has an itinerant $d$-band which is allowed to hybridize with the $f$-electrons; these two layers hybridize with each other through a hopping term, $t_z$, as shown in Fig.\,\ref{fig:bilayer}.
As a first step towards a better understanding of the nature of such an interface, it is convenient
to resort to less restrictive methodologies.
Thus, we use an unrestricted Hartree-Fock approximation \cite{Costa17a,Costa18a} to investigate the properties of this system.
Despite working on a two-dimensional lattice, our results should serve as a qualitative guide both to the interplay between opposing tendencies in the ground state, and to a three-dimensional construction, with the two layers as the repeating unit.
The layout of the paper is as follows. 
In Sec.\,\ref{sec:model} we present the full Hamiltonian and its Hartree-Fock version  (with the details of derivation being left to the Appendix). 
Section \ref{sec:zerot} presents the results for the ground state transitions, while Sec.\,\ref{sec:finitet} is concerned with effects of temperature. 
And, finally, Sec.\,\ref{sec:concl} summarizes our findings.

\section{Model and methods}
\label{sec:model}

The system is described by a two-layer Hamiltonian, 
\begin{align}
\mathcal{H} =&-t\sum_{\langle \mathbf{i,j}\rangle\!,\sigma}\left(c_{\mathbf{i}\sigma}^\dagger c_{\mathbf{j}\sigma}^{\phantom{\dagger}}+d_{\mathbf{i}\sigma}^\dagger d_{\mathbf{j}\sigma}^{\phantom{\dagger}}+\mathrm{h.c.}\right) -U\sum_{\mathbf{i}}n^{c}_{\mathbf{i}\uparrow}n^{c}_{\mathbf{i}\downarrow}\nonumber \\
&-t_{z}\sum_{ \mathbf{i}\!,\sigma}\left(c_{\mathbf{i}\sigma}^\dagger d_{\mathbf{i}\sigma}^{\phantom{\dagger}}+\mathrm{h.c.}\right) 
+J\sum_{\mathbf{i}} \mathbf{s}^{d}_\mathbf{i}\cdot \mathbf{S}^{f}_{\mathbf{i}}.
\label{eq:dEHM}
\end{align}
The first term describes the hopping of electrons both on the top (conduction, $c$)  and bottom ($d$) layers (see Fig.\,\ref{fig:bilayer}), with  $c _{\mathbf{i}\sigma}^{\phantom{\dagger}}$($c ^\dagger_{\mathbf{i}\sigma}$) and $d _{\mathbf{i}\sigma}^{\phantom{\dagger}}$($d ^\dagger_{\mathbf{i}\sigma}$) denoting the respective annihilation (creation) operators in standard second quantization formalism; h.c.\ stands for hermitian conjugate of the previous expression.
The hopping integral, $t$, sets the energy scale and is assumed to be the same on both layers, with $\langle \mathbf{i,j}\rangle$ denoting nearest neighbor sites on the same layer, and $\sigma=\,\uparrow,\downarrow$ standing for the electron spins states; we also set the Boltzmann constant, $k_\text{B}$, to unity.
The second term favors pairing, driven by an attractive on-site coupling \cite{Micnas90}, $-U<0$, solely on the (top) $c$-layer, with $n^{c}_\mathbf{i}$ being the $c$-orbital number operator on site $\mathbf{i}$; a one-dimensional model with Kondo and pairing interactions was considered in Ref.\,\cite{Bertussi09}. 
The third term is the interlayer hopping along the vertical direction, so that $t_z$ effectively describes the degree of hybridization between $c$ and $d$-orbitals. 
And, finally, the fourth term corresponds to the Kondo exchange coupling (strength $J$)  between the $d$-electrons on the bottom layer and the local $f$-moments.

It is important to notice that in the noninteracting limit $(U=J=0)$, and at half filling, the system is metallic up to $t_{z}=4t$, beyond which a band gap opens. 
Therefore, as we tune in $U$, $J$, $t_z$ ($<4t$), and the band filling competing effects may emerge between superconducting (SC), antiferromagnectic (AFM), and singlet (Kondo) phases. 
In order to investigate the main features of this competition, we adopt a Hartree-Fock approach \cite{Costa18a} which allows us to probe the occurrence of spiral magnetic phases characteristic of the KLM \cite{Costa17a}. 
To this end, we write the spin operators for the $d$ and $f$-orbitals in a fermionic basis as 
\begin{equation}
\mathbf{s}_{\textbf{i}}^{d} = \dfrac{1}{2}  \sum_{\alpha , \beta=\pm}d_{\textbf{i}\alpha}^{\dagger} \pmb{\sigma}_{\alpha , \beta} d_{\textbf{i}\beta}^{\phantom{\dagger}},
\label{fermi1}
\end{equation}
and
\begin{equation}
\mathbf{S}_{\textbf{i}}^{f} = \dfrac{1}{2}  \sum_{\alpha , \beta = \pm }f_{\textbf{i}\alpha}^{\dagger} \pmb{\sigma}_{\alpha , \beta} f_{\textbf{i}\beta}^{\phantom{\dagger}},
\label{fermi2}
\end{equation}
with $\pmb{\sigma}_{\alpha , \beta}$ denoting the elements of the Pauli matrices, and $f _{\mathbf{i}\sigma}^{\phantom{\dagger}}$($f ^\dagger_{\mathbf{i}\sigma}$) being creation (annihilation) operators for localized electrons. 
With the transformations of Eqs.\,\eqref{fermi1} and \eqref{fermi2}, the Hamiltonian of Eq.\,\eqref{eq:dEHM} acquires quartic operators for the Kondo terms, in addition to those for the Hubbard one.
Following the procedure outlined in the Appendix, we use a Hartree-Fock approximation to write the Hamiltonian in a quadratic form, which becomes
\begin{widetext}
\begin{align}
\mathcal{H}_\text{HF} &=
\sum_{\kv\sigma}\bigg[ (\varepsilon_{\kv}- \mu)(c_{\kv\sigma}^{\dagger}c_{\kv\sigma}+d_{\kv\sigma}^{\dagger}d_{\kv\sigma})-t_{z}(d_{\kv\sigma}^{\dagger} c_{\kv\sigma}+ h.c.)+
 \dfrac{3JV}{4}(d_{\kv\sigma}^{\dagger} f_{\kv\sigma} + h.c)+\varepsilon_{f}f_{\kv\sigma}^{\dagger}f_{\kv\sigma}\bigg]\nonumber\\
& +\sum_{\kv}\bigg[-U(P_{cc} c^{\dagger}_{\kv\uparrow}c^{\dagger}_{-\kv\downarrow}+h.c.)+\dfrac{J}{4}(P_{df} f^{\dagger}_{\kv\uparrow}d^{\dagger}_{-\kv\downarrow}+P_{df}f^{\dagger}_{\kv\downarrow}d^{\dagger}_{-\kv\uparrow}+h.c.)+\dfrac{J}{2}(m_{f}d_{\kv\uparrow}^{\dagger} d_{\textbf{k+Q}\downarrow}\nonumber\\
& -m_{d}f_{\kv\uparrow}^{\dagger} f_{\textbf{k+Q}\downarrow}+h.c.) \bigg]
+ UNP_{cc}^2+JN\bigg( m^{d}m^{f} +\dfrac{3V^{2}}{2} -\dfrac{P_{df}^2}{2} \bigg)+ N(2\mu n - \varepsilon_{f}n_{f}),
\label{eq:HHF}
\end{align}
\end{widetext}
at the price of introducing mean-field variables that should be determined self-consistently, as described below. Here, we include $\mu$ and $\varepsilon_f$ as Lagrange multipliers, in order to fix (on average) the electronic density to the desired values in their respective bands, and $\varepsilon_\kv=-2t(\cos k_x+ \cos k_y)$ is the dispersion for the bare conduction electrons.
Notice that the electronic density of the conduction bands $c$ and $d$ is fixed on average as \linebreak$n=(n_c+n_d)/2 $, and for the localized electrons it is set as $n_f=1$; $N$ is the number of sites. 

A quantitative measure of the hybridization between $d$ and $f$ bands is given by
\begin{equation}
V_{\textbf{i}} = \dfrac{1}{4} \sum_{\sigma , \sigma' = \pm }\langle d_{\textbf{i}\sigma}^{\dagger} \mathbb{1}_{\sigma \sigma'}^{\phantom{\dagger}} f_{\textbf{i}\sigma'}^{\phantom{\dagger}}+\text{h.c.}\rangle,
\label{hybrid}
\end{equation}
where $\mathbb{1}$ is the unit operator, 
and we assume $V_\mathbf{i}=V,\,\forall\mathbf{i}$
\footnote{We have checked the contribution of triplet hybridization terms \cite{Costa17a,Costa18a}, and found that they are negligible.}.
The nature of magnetic ordering enters in $\mathcal{H}_\text{HF}$ through the wavevector $\mathbf{Q}$ and the magnetization amplitudes, $m_d$ and $m_f$; these describe the average of $d$ and $f$ spins operators as
\begin{equation}
\langle\mathbf{s}_{\textbf{i}}^{d} \rangle= 
- m_{d} \big[
\cos(\textbf{Q}\cdot \textbf{R}_\mathbf{i}), \sin(\textbf{Q}\cdot \textbf{R}_\mathbf{i}), 0
\big]~,
\label{ansatz1}
\end{equation}
and
\begin{equation}
\langle\mathbf{S}_{\textbf{i}}^{f}\rangle =
m_{f} \big[
\cos(\textbf{Q}\cdot \textbf{R}_\mathbf{i}), \sin(\textbf{Q}\cdot \textbf{R}_\mathbf{i}), 0
\big]~,
\label{ansatz2}
\end{equation}
where $\mathbf{R}_\mathbf{i} $ is the position vector of site \textbf{i} on the lattice. 

And, finally, we note that different dimensionless pairing amplitudes appear in $\mathcal{H}_\text{HF}$, defined as 
\begin{equation}
	P_{\alpha\beta}\equiv \langle \alpha^\dagger_{\iv \uparrow}\beta^\dagger_{\iv\downarrow} \rangle = \langle \beta_{\iv\downarrow} \alpha_{\iv \uparrow} \rangle, 
\label{eq:Deltaab}	
\end{equation}	
where $\alpha,\beta=c,d,f $. 
They measure the relative importance of the different pairing channels, and 
we recall that $P_{cc}$ connects with the Hartree-Fock superconducting gap in the single layer attractive Hubbard model through $\Delta = 2|U|\,P_{cc}$.

The mean-field Hamiltonian, written in a basis of spinors with 12 components, can be diagonalized to obtain the eingenvalues $\lambda_{\mathbf{k},\nu}$.
Then, the Helmholtz free energy is written as
\begin{equation}
F=-\frac{1}{2} k_BT\sum_\mathbf{k}\sum_{\nu=1}^{12}\ln\bigg[ 1+ \exp\bigg(-\frac{\lambda_{\mathbf{k},\nu}}{k_BT} \bigg) \bigg],
\label{hfe}
\end{equation}
apart from a constant term.
Given this, the mean-field variables are then determined self-consistently by minimizing the Helmholtz free energy with aid of the Hellmann-Feynman theorem, 
\begin{equation}
\begin{split}
 & \Bigg\langle \frac{\partial F}{\partial \mu} \Bigg\rangle =\Bigg\langle \frac{\partial F}{\partial \epsilon_{f}} \Bigg\rangle =\Bigg\langle \frac{\partial F}{\partial Q_{\alpha}} \Bigg\rangle =\Bigg\langle \frac{\partial F}{\partial m_{f}} \Bigg\rangle=\Bigg\langle \frac{\partial F}{\partial m_{d}} \Bigg\rangle=\\&=  \Bigg\langle \frac{\partial F}{\partial V} \Bigg\rangle  =\Bigg\langle \frac{\partial F}{\partial P_{cc}} \Bigg\rangle =\Bigg\langle \frac{\partial F}{\partial P_{df}} \Bigg\rangle  =0.
\end{split}
\label{eq:HFthm}
\end{equation}
The calculations have been carried out for lattices of size $N_x \times N_y = 200 \times 200$ with periodic boundary conditions.

Once convergence is achieved, we are able to investigate spectral properties, by calculating the single-particle Green's function (see, e.g.\ Refs.\ \cite{nandini,Beach} for details),
\begin{equation}
G^\alpha_\sigma(\mathbf{k},\tau)=\langle \mathcal{T} [\alpha_{\mathbf{k},\sigma}^{\phantom{\dagger}}(0) \alpha^\dagger_{\mathbf{k},\sigma}(\tau)]\rangle, 
\end{equation}
where $\mathcal{T}$ is the time ordering operator, $\tau $ is the time, and $\alpha =c,d,f$. 
Taking the time Fourier transform, and performing an analytical continuation allow one to obtain the spectral function,
\begin{equation}
A^\alpha_\sigma(\mathbf{k},\omega)=\dfrac{1}{\pi}\operatorname{Im} G^\alpha_\sigma(\mathbf{k},\omega+ i 0^+),
\end{equation}
from which we obtain the density of states (DOS) as
\begin{equation}
D^\alpha (\omega)=\sum_{\mathbf{k},\sigma}A^\alpha_\sigma(\mathbf{k},\omega).
\end{equation}

\color{black}

\section{GROUND STATE PROPERTIES}
\label{sec:zerot}

Let us start by examining the ground state properties. 
We first discuss (Sec.\,\ref{ssec:AFMK}) how the presence of a conducting layer without pairing (that is, we set $U=0$ on the conduction layer) affects the magnetic transition between an
AFM state and the Kondo singlet state. 
Then, in Sec.\,\ref{ssec:TransportK} we concentrate on the Kondo phase and study the influence of an extra conducting layer (still without pairing) on the the different conducting channels.    
Finally, in Sec.\,\ref{ssec:Uneq0} we
switch on the attractive interaction on the conduction layer, and analyze how the transport properties change.  

\subsection{The AFM-Kondo transition ($U=0$)}
\label{ssec:AFMK}

\begin{figure}[t]
\centering
\includegraphics[scale=0.3]{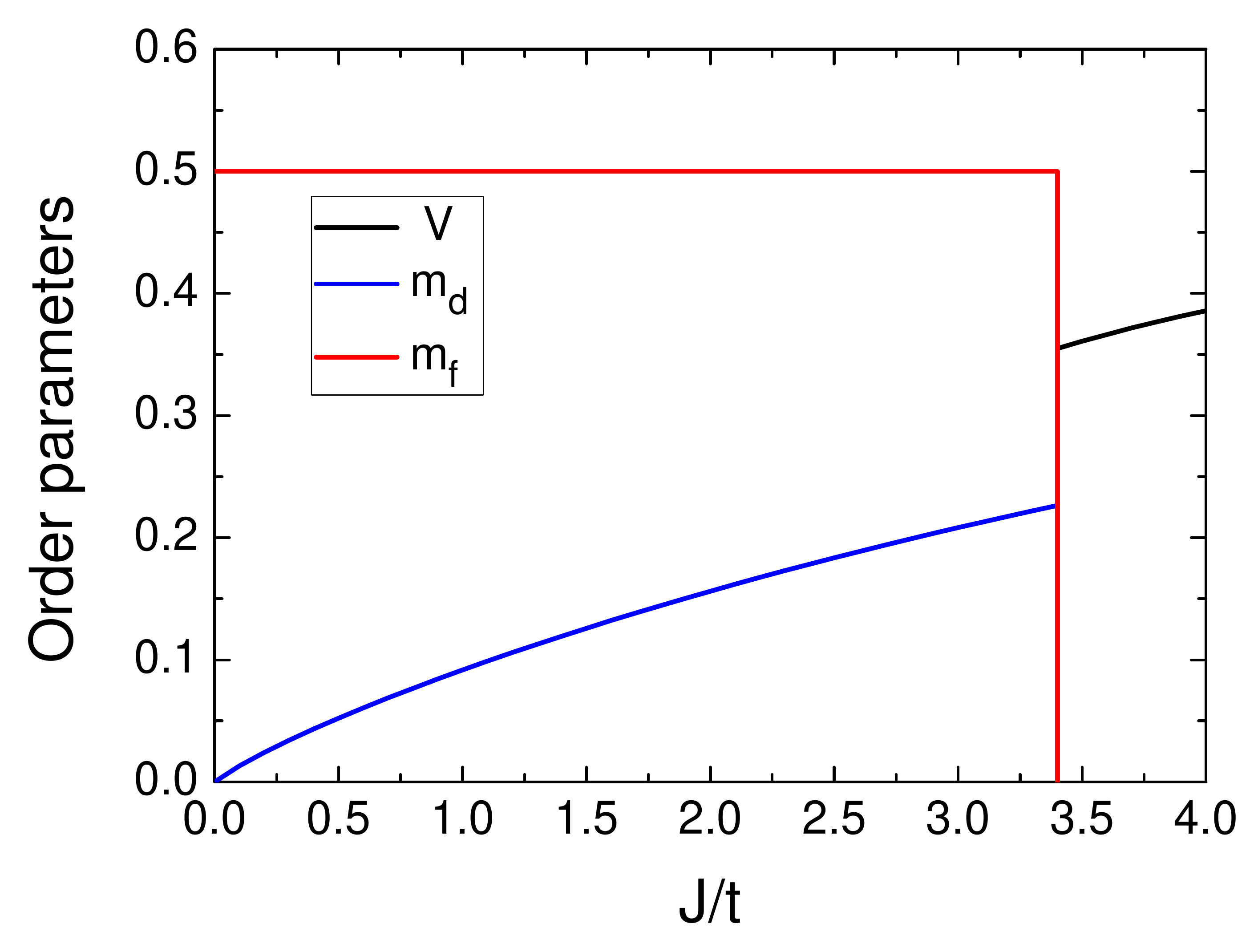} 
\caption{(Color online) Hybridization and magnetization amplitudes as functions of the Kondo coupling, $J/t$. 
All data are for $n=1.0$ and $t_z=1.0$.
}
\label{fig:opxj}
\end{figure}

\begin{figure}
\centering
\includegraphics[scale=0.3]{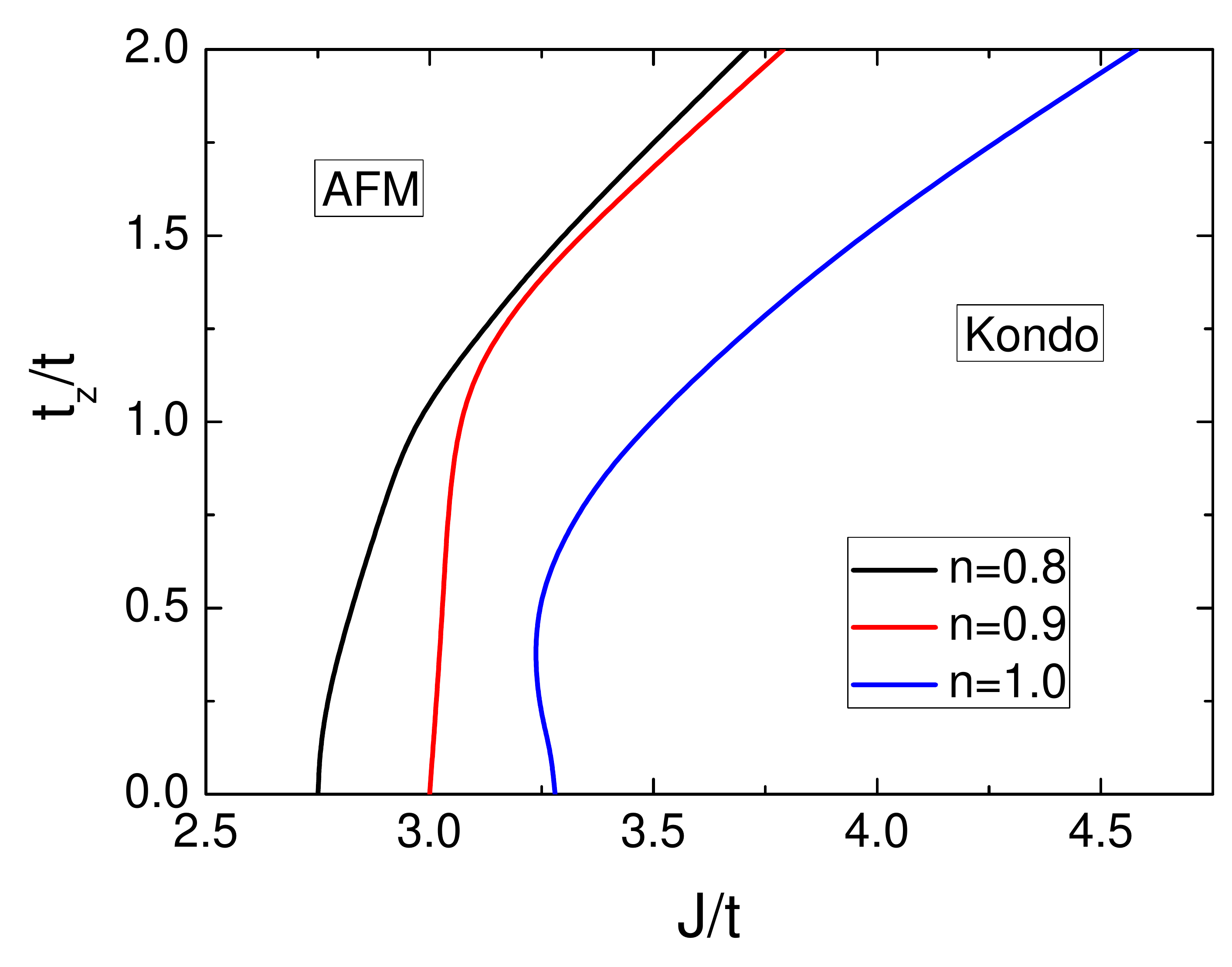} 
\caption{(Color online) $t_z\times J$ phase diagrams for the Kondo-antiferromagnetic (AFM) phase transition at different band fillings, when $U=0$.}
\label{fig:tzxj}
\end{figure}

The free energy is minimized for fixed electron density, $n$, and Kondo coupling $J$; at first we set $t_z/t=1$. 
From this we obtain the magnetization amplitudes, $m_d$ and $m_f$, as well as the hybridization, $V$; the results for half filling (i.e.\ one local moment and one electron per site, $n=1$) are shown in Fig.\,\ref{fig:opxj}.
Similarly to the single-layer KLM \cite{Costa17a}, there is a critical value of the coupling, $J_c$, which, at half filling, separates a N\'eel phase of the $f$-electrons (whose interaction is mediated by the $d$-electrons) from a Kondo-singlet phase. 
The appearance of the Kondo phase is accompanied by $V$ jumping from zero to a finite value.
One should notice, though, that in the present case the critical coupling $J_c$ is larger than its value for the single-layer KLM. Indeed, as suggested for the KLM \cite{Sen16,Peters2016} and for the PAM~\cite{Hu2017}, an additional metallic layer may enhance the effective RKKY interaction between the local moments,
as a result of the induced magnetic order in the conduction layer; simultaneously, the interlayer hopping, $t_z$, weakens Kondo-singlet formation. 
The latter effect can be seen more clearly if one allows $t_z/t$ to vary: as shown in Fig.\,\ref{fig:tzxj}, a Kondo-singlet phase gives way to an AFM state  above some critical value of $t_z$, which depends on $J$.

\begin{figure}
\centering
\includegraphics[scale=0.45]{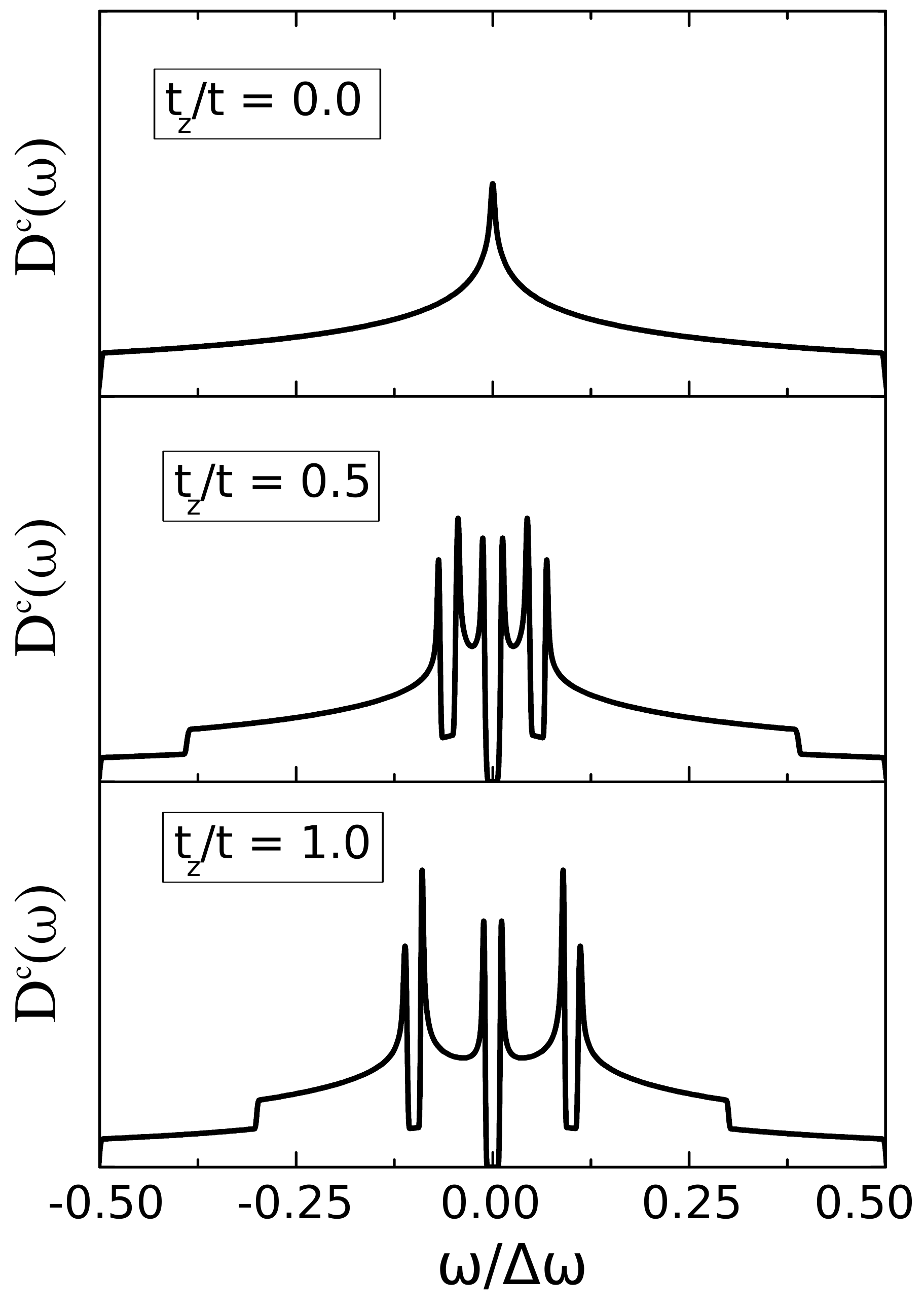} 
\caption{(Color online) Density of states for the $c$-band as a function of $\omega/\Delta\omega$ (where $\Delta\omega=\omega_{max}-\omega_{min}$), for different values of the interplane hybridization, $t_z$, and for $J/t=0.8$, deep in the AFM phase: 
The low energy spin excitations cause a gap opening even in the metallic layer.}
\label{fig:dcw}
\end{figure}

It is also interesting to examine the effect of $t_z$ on the $c$-band, deep in the antiferromagnetic phase. 
Figure \ref{fig:dcw} shows the $c$-band DOS for three different values of $t_z$, from which we see that a gap opens for $t_z\neq 0$. 
This occurs as a result of $c$-electrons acquiring an induced polarization, so that they also take part in the RKKY mechanism responsible for the antiferromagnetic ordering of $f$-electrons.

Away from half filling the system is no longer an insulator, so that the increased electron mobility  weakens the RKKY interaction. 
Accordingly, Fig.\,\ref{fig:tzxj} shows that as $n$ decreases from half filling, $J_c$ for the singlet phase decreases. 
Within the mean-field approach used here [see Eqs.\,\eqref{ansatz1} and \eqref{ansatz2}], which allows for a spiral-type alignment of the spins, the magnetic wavevector  minimizing the energy is $\mathbf{Q}=(\pi,q)$ [or its degenerate pair, $(q,\pi)$]. 
In the range $0.7 \leq n\leq 1$, $q$ does not change significantly with $J$, although for fixed $J<J_c$ we found that $q\to \pi$ monotonically when $n\to 1$.


Further decrease in the electronic density drives the system to a ferromagnetic state, as in the single-layer KLM. However, since our main interest here is in the effect of the extra layer on the transport properties in the Kondo-singlet phase, we have not pursued a detailed analysis of the magnetic properties.

\subsection{Transport properties in the Kondo phase ($U=0$)}
\label{ssec:TransportK}
\begin{figure*}[t]
\centering
\includegraphics[scale=0.57]{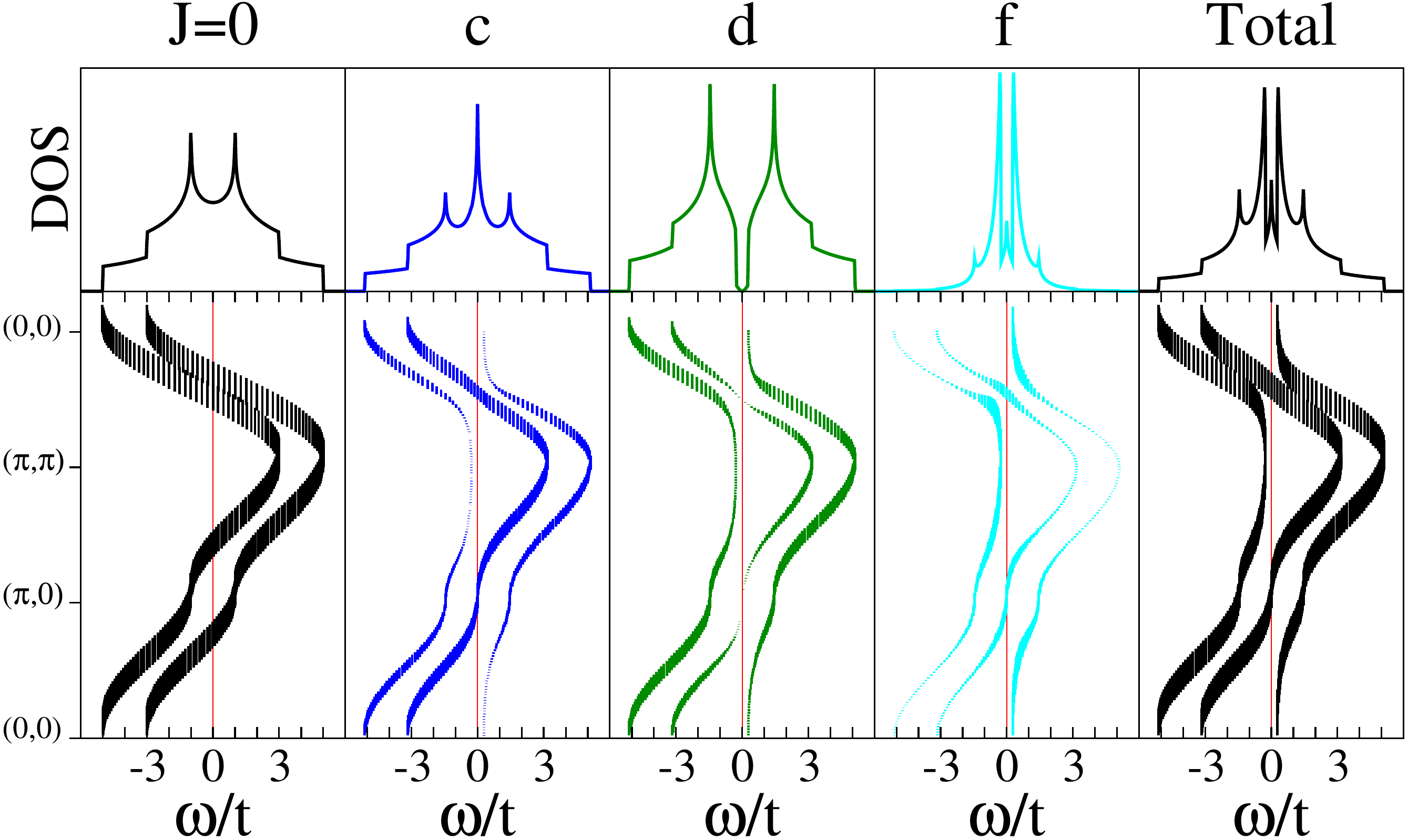} 
\caption{(Color online) Density of states (DOS; top row) and spectral function $A(\mathbf{k},\omega)$ (bottom row) for the bilayer with $U=0$ and $t_z/t=1.0$. 
In succession, from left to right the columns show: data for decoupled $d$ and $f$ bands ($J=0$), and data in the Kondo phase $(J/t=3.75)$ resolved for the $c$- $d$- and $f$-bands, as well as total data. 
All data are plotted as functions of energy, $\omega/t$ (in units of $\hbar^{-1}$), which is set to zero at the Fermi energy of the half filled band.}
\label{DOS}
\end{figure*}

Let us then examine the effects of the $c$-layer on the transport properties within the Kondo phase. Figure \ref{DOS} shows the results for both the density of states and the spectral function for each band. The discrete nature of the spectrum leads to a sequence of $\delta$-functions (shown as spikes) representing  $A^\alpha_\sigma(\mathbf{k},\omega)$, the magnitude of which being associated with the height of the spikes; in addition, the spikes trace out $\omega(\mathbf{k})$, the Hartree-Fock energy bands. The leftmost panel shows $D(\omega)$ and $A^\alpha_\sigma(\mathbf{k},\omega)$ for the decoupled $d$ and $f$-bands, i.e.\ $J=0$. 
We see that the hopping, $t_z$, between the $d$ and $c$-bands splits the two degenerate square lattice DOS's, a feature also present in the spectral function.
Nonetheless, the value $t_z=t$ used in this case is not enough to split the bands to the point of generating an insulating state, which would happen for $t_z>4t$.

Let us now switch on the coupling $J$ in such way that the system is in the Kondo (singlet) phase for half filling and $t_z=t$. 
The top row of Fig.\,\ref{DOS} shows the density of states for the $c$-, $d$-, and $f$-orbitals.  
We see that the $c$-band preserves its metallic character while the other bands drastically change the behavior in comparison with the single-layer KLM \cite{Beach}.  
First, the usual finite gap in the $d$-band (usually referred to as the $c$-band in the single-layer KLM) here becomes a pseudogap, in the sense that the density of states only vanishes at this single, isolated energy. 
Second, the $f$-band, which also displays a gap in the single layer KLM, here not only acquires a metallic character, but the DOS is peaked exactly at the Fermi energy for half filling.   
Finally, these features manifest themselves in the total density of states: instead of a gap, as in the single-layer KLM \cite{Beach}, the overall metallic character of the system is evident from the peak at the Fermi energy, as shown in the rightmost column in Fig.\,\ref{DOS}; such a feature is explored below. 

The bottom row of Fig.\,\ref{DOS} shows the spectral function for each band. 
We see clearly that the Kondo coupling causes yet another splitting of the bands; this, in turn, is accompanied by the appearance of nearly dispersionless regimes which favor the formation of nearly localized states.  
However, while this tendency is compensated by non-zero spectral weight at the Fermi energy for both $c$- and $f$-bands, there is no spectral weight at the Fermi energy for the $d$-band, thus giving rise to the pseudogap.
Indeed, from the procedure outlined in the Appendix, we may extract the contributions at the Fermi wavevectors, $\mathbf{k}_\text{F} \approx (\pi,0)$ and $(\nicefrac{\pi}{2},\nicefrac{\pi}{2})$, as
\begin{align}
	A^c_\sigma({\mathbf{k}_F},0) &=  \dfrac{9J^2V^2}{9J^2V^2 + 16t_z^2 }\\
 	A^d_\sigma({\mathbf{k}_F},0)&=  0\\
 	A^f_\sigma({\mathbf{k}_F},0) &=  \dfrac{16t_z^2}{9J^2V^2 + 16t_z^2 }.
\end{align}

\begin{figure}[t]
\centering
\includegraphics[scale=0.36]{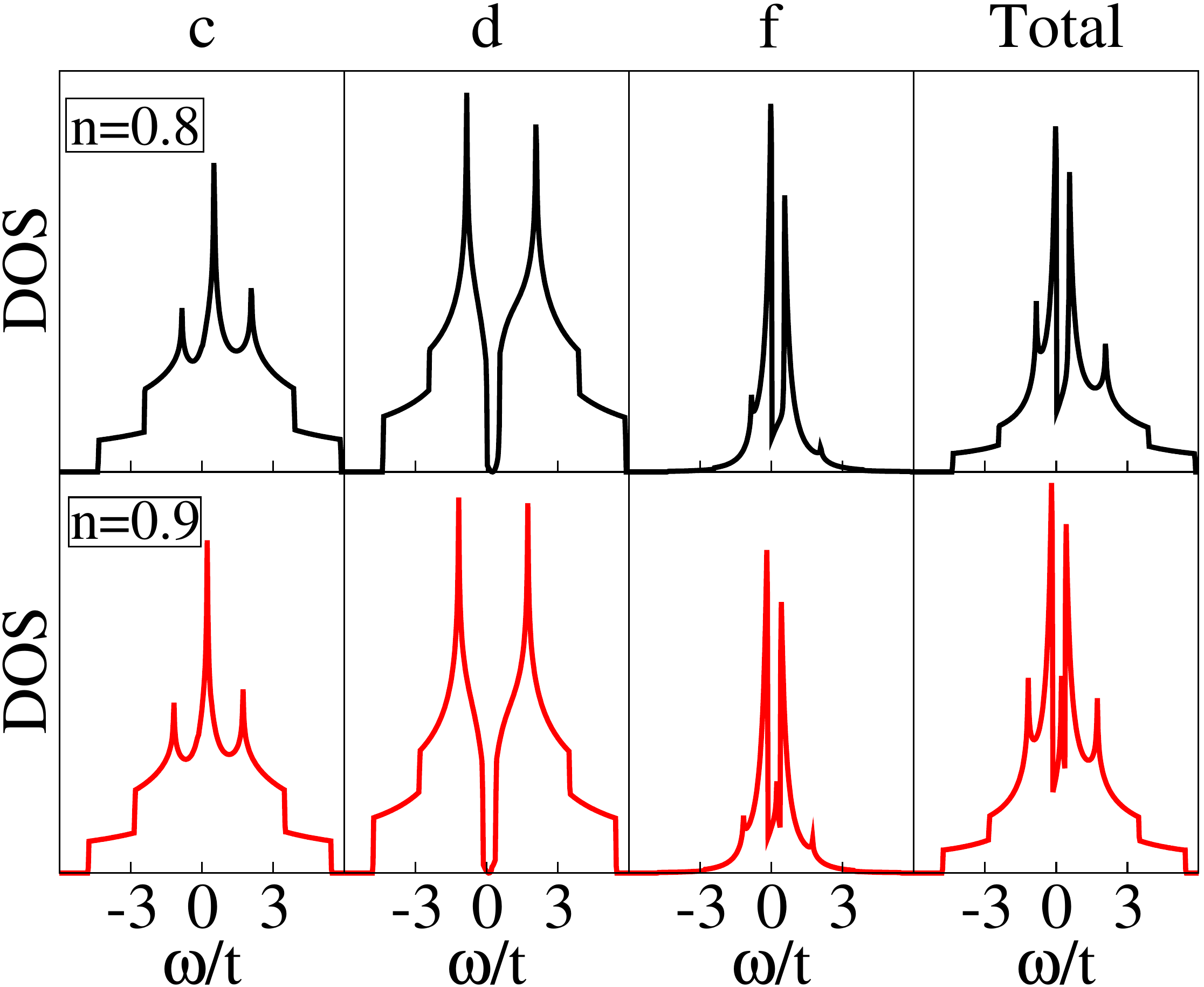} 
\caption{
(Color online) Same as the top row of Fig.\,\ref{DOS}, but for $n=0.8$ (top row) and $n=0.9$ (bottom row). The Fermi energy is set at the origin in both cases.
}
\label{fig:DOS-0908}
\end{figure}

Away from half filling, our self-consistent procedure leads to quasi-rigid bands, whose corresponding DOS's are shown in Fig.\,\ref{fig:DOS-0908}. 
The most significant difference relative to the $n=1$ case is the suppression of the peak at the Fermi energy for the $f$-electrons, which is also manifested in the total DOS. 
Nonetheless, the metallic character is preserved.
In this respect, we recall that the single-layer KLM displays a gap in the Kondo phase at half filling which (within Hartree-Fock) is approximately rigidly displaced upon doping, leading to a metallic phase~\cite{Beach}.

\begin{figure}[t]
\centering
\includegraphics[scale=0.3]{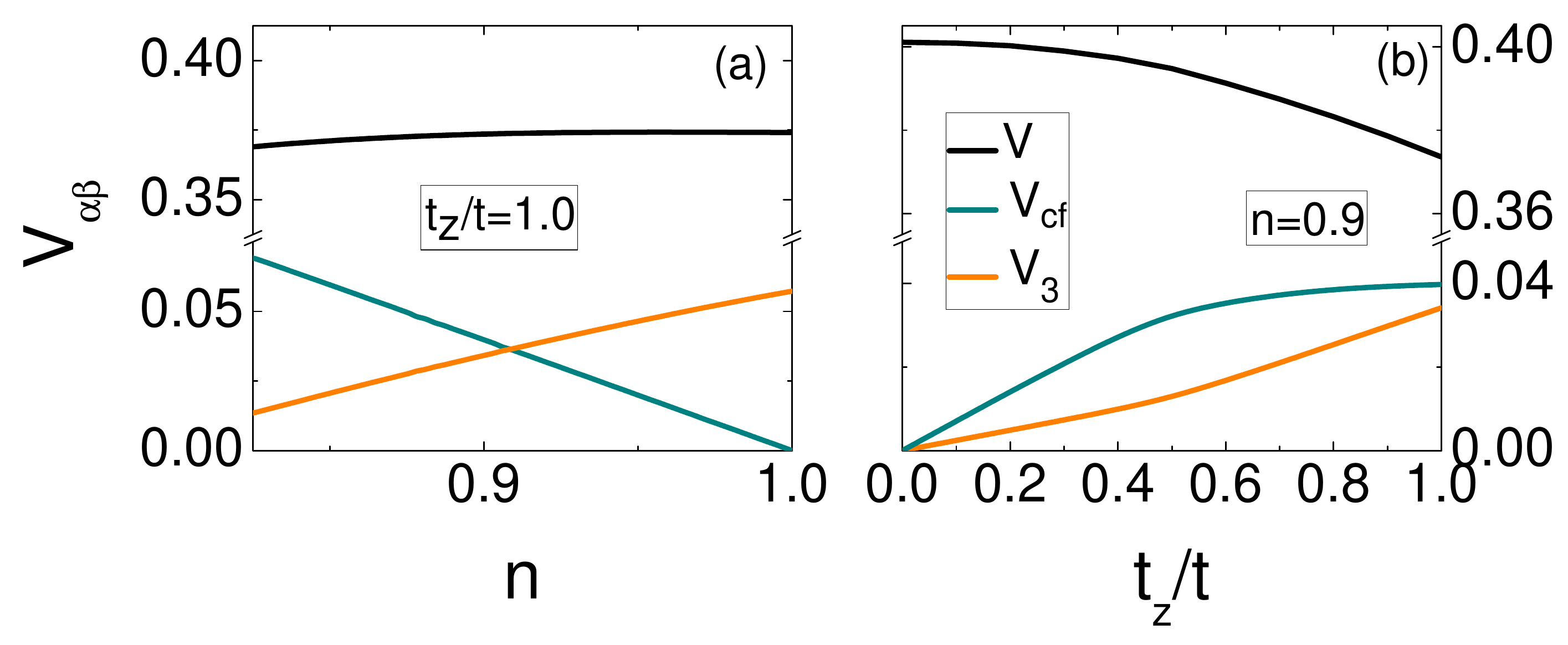} 
\caption{(Color online) Calculated hybridizations: $V$ ($d$-$f$), $V_{cf}$, and  $c$-$d$-$f$  $V_3$: (a) as functions of the band filling, for $t_z=t$, and (b) as functions of $t_z/t$, for $n=0.9$. In both cases, $J/t=3.75$.
}
\label{fig:Vab}
\end{figure}

Further insight into how the presence of an extra metallic layer affects the otherwise insulating character, can be obtained by defining 
\begin{align}
	V_{cf} &=  \dfrac{1}{2} \langle c_{\textbf{i}\sigma}^{\dagger} f_{\textbf{i}\sigma}^{\phantom{\dagger}}+\text{h.c.}\rangle,\\
	V_{3} &=  \dfrac{1}{2} \langle c_{\textbf{i}\sigma}^{\dagger} d_{\textbf{i}\sigma}^{\phantom{\dagger}} d_{\textbf{i}\sigma}^\dagger f_{\textbf{i}\sigma}^{\phantom{\dagger}}+\text{h.c.}\rangle,
\end{align}	
which respectively probe $c$-$f$ hybridization and the triple  $c$-$d$-$f$ hybridization.
When varying the filling, as displayed in Fig.\,\ref{fig:Vab}\,(a), the $d$-$f$ hybridization, $V$, dominates (in the range of $n$ considered), but the relative importance of $V_{cf}$ and $V_3$ drastically changes across this density range. 
Indeed, we see that at half filling there is no direct $c$-$f$ hybridization, but the $c$-$d$-$f$ hybridization is maximum. 
The emergence of a pseudogap in the $d$ channel may then be attributed to the singlet resonating between $d$-$f$ and $c$-$d$ combinations, which enhances the mobility of the $f$-electrons while localizing $d$-electrons.
For $n=0.9$, Fig.\,\ref{fig:Vab}\,(a) shows that $V_{cf}>V_3$ so that resonating singlets are less likely. 
Further, Fig.\,\ref{fig:Vab}\,(b) shows that these additional hybridization features decrease as $t_z$ decreases, as one may expect.

\subsection{Pairing in the Kondo phase}
\label{ssec:Uneq0}

Before proceeding with the effects caused by having $|U|$ on the additional metallic layer, it is worth recalling what happens in the case of a single-layer KLM with an attractive interaction in the $d$-band, in the regime where $d$ and $f$-electrons are strongly hybridized into singlets~\cite{Costa18a}.
In this case, the tendency for local pairing in the $d$-band requires breaking the local $d$-$f$ singlets and, consequently, coexistence between Kondo phase and ($d$-band) superconductivity is unlikely. Indeed, due to the high energetic cost to break the singlets, one needs a very strong attractive interaction to generate pair coherence~\footnote{See, e.g., Ref.\,\cite{Costa18a} for the ground state phase diagram on a simple cubic lattice, which, within Hartree-Fock, is qualitatively similar to what happens on a square lattice.}.

By contrast, in the bilayer system the pairing interaction acts solely on the $c$-electrons of the extra layer.
Therefore, for any small, finite $|U|$ pairs can be formed on the $c$-layer, without the need to break up singlet pairs formed in the $d$-$f$ layer. 
However, as discussed earlier, the inclusion of such metallic layer changes the spectral weight at the Fermi level, which may lead to new pairing features.
In view of this, it is interesting to examine
the behavior of the pairing amplitudes defined in Eq.\,\eqref{eq:Deltaab}, for both intra- and inter-orbital channels. 

\begin{figure}[t]
\centering
\includegraphics[scale=0.3]{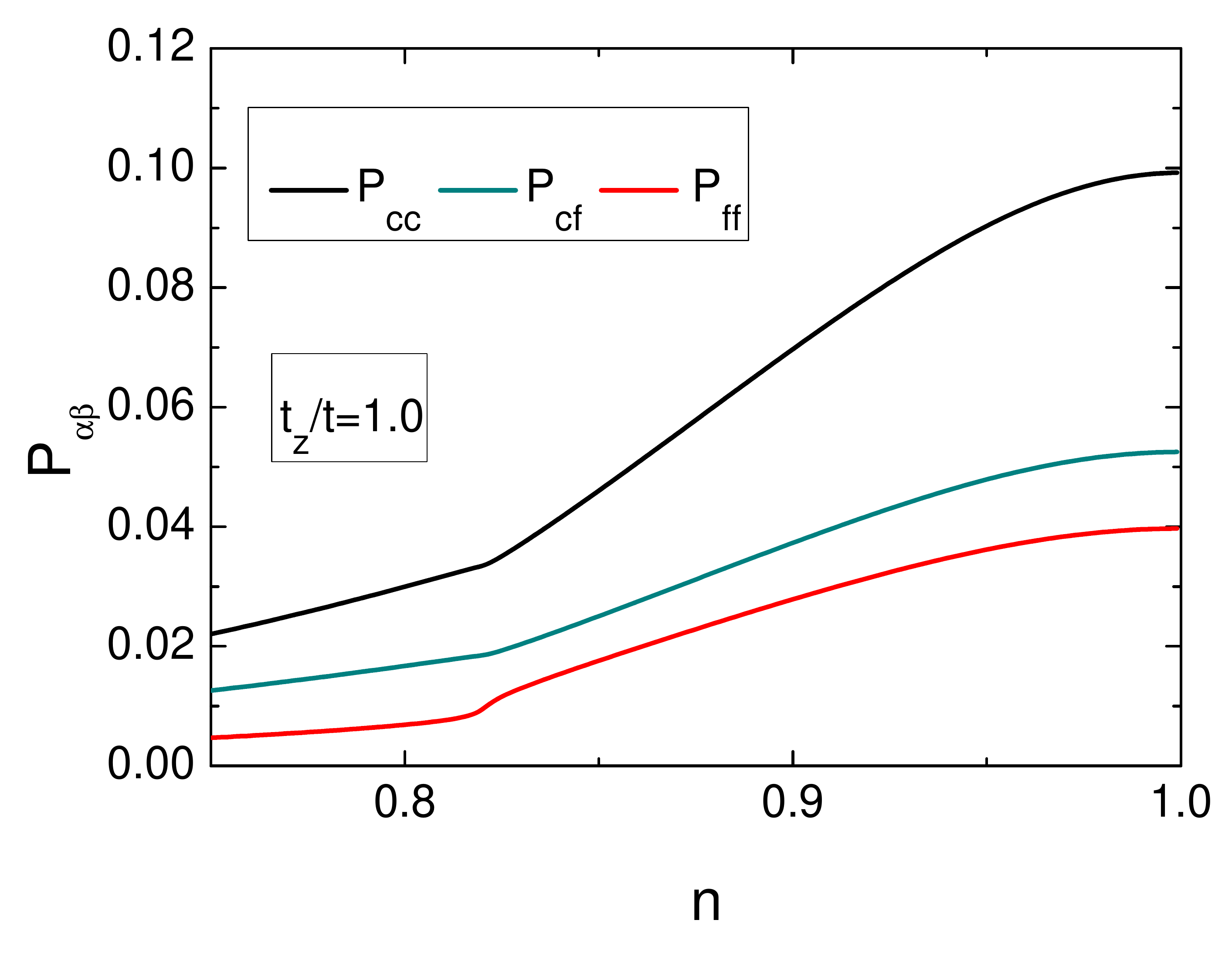} 
\caption{(Color on line) Pairing amplitudes as functions of electronic density for $J/t=4.0$ and $U/t=2.0$.
}
\label{dxnc}
\end{figure}

Figure \ref{dxnc} displays the pairing amplitudes as functions of the band filling, for fixed $J/t$, $U/t$, and $t_z/t$. 
The dominant channel in the range of fillings considered corresponds to having the paired electrons in $c$-orbitals, which is to be expected since the pairing interaction is confined to the $c$-layer.
Interestingly, as also displayed in Fig.\,\ref{dxnc}, other pairing contributions appear:
the second dominant pairing amplitude involves $c$ and $f$-electrons, followed by the $ff$ one.
Notice that these were the two orbitals with finite spectral weight at the Fermi level, for the $U=0$ case.
By contrast, other possibilities involving the $d$-orbital are always suppressed, which may be attributed to the presence of the pseudogap in the $d$-DOS.
Furthermore, the analysis of the DOS (not shown) provides a unique SC gap $\Delta$ for all orbitals; that is, the entire system is superconducting, not only the additional layer.
Thus, recalling that $\Delta/t \sim P$ (for a single-orbital approach), these results suggest that we have a \textit{superconducting Kondo phase}, i.e.\ one in which pairing occurs amongst the quasiparticles of the Kondo fluid formed by the $c$-, $d$-, and $f$-electrons.

\begin{figure}
\centering
\includegraphics[scale=0.3]{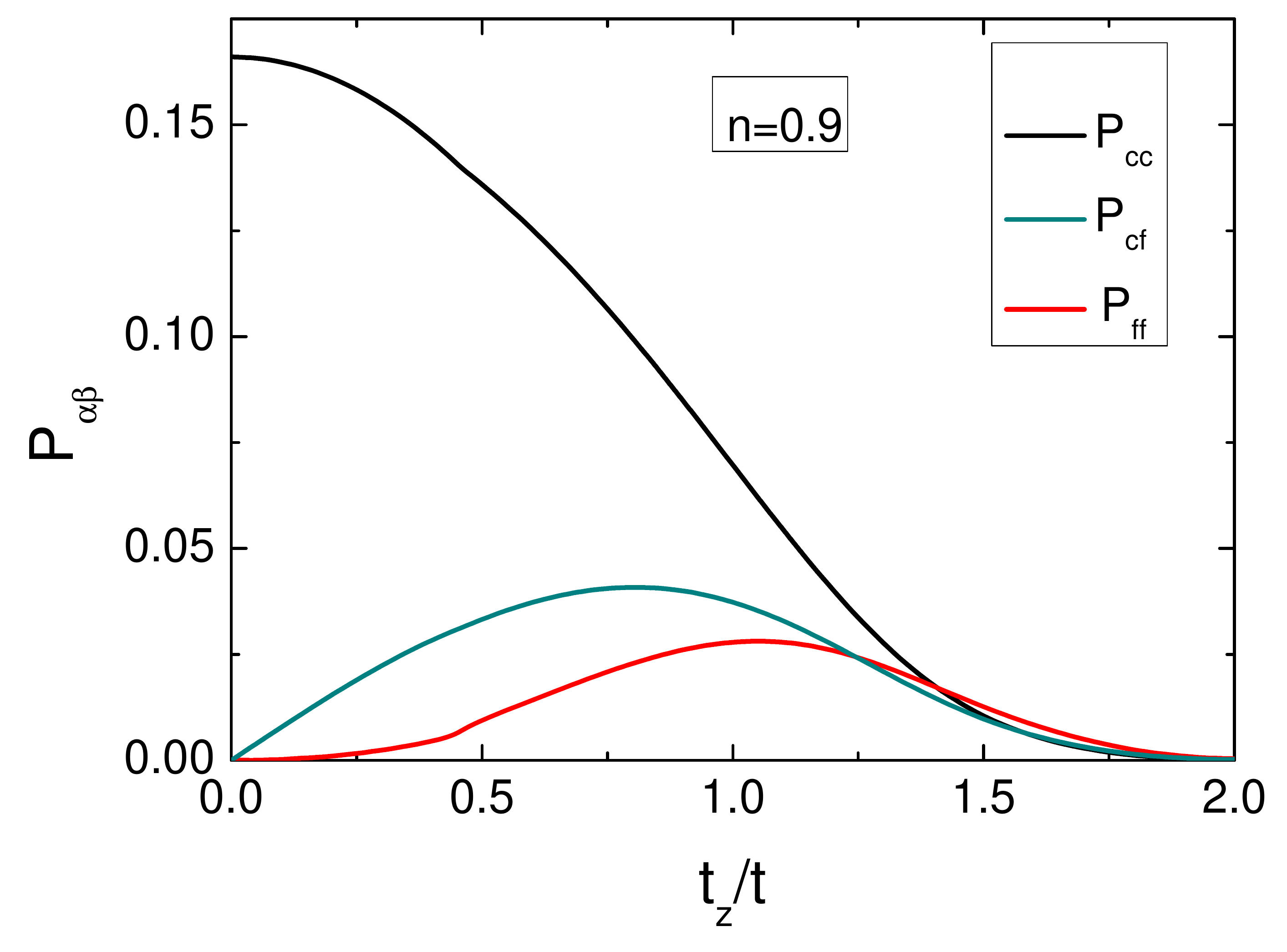} 
\caption{(Color on line) Pairing amplitudes in three different channels as functions of the interlayer hopping, $t_z/t$, for $J/t=4.0$, $U/t=2.0$ and $n=0.9$.
}
\label{dxtz}
\end{figure}

Further insight into this interplay between pairing in different channels can be gained by examining the electron densities on $c$ and $d$-orbitals. 
At half filling, the minimum free energy corresponds to $ n_c=n_d=1$, but when $n=(n_c+n_d)/2<1$ the Hubbard attraction induces electron migration to the $c$-layer, thus causing charge imbalance, $n_c>n_d$. 
This migration, in turn, depends on the hybridization between $c$ and $d$-orbitals. 
Figure \ref{dxtz} shows the pairing amplitudes in different channels as functions of the interlayer hopping, $t_z$, for fixed $J/t$, $U/t$, and $n$.
The increasing $t_z$ hybridization is deleterious to $cc$ pairing,
while it does not lead to dominant pairing from the remaining channels; the overall effect of large $t_z$ is therefore to suppress superconductivity altogether. 
Similar behavior occurs for other fillings within the Kondo phase. We conclude that the emergence of this superconducting state was only made possible due to the spatial separation between the $c$ and $d$-orbitals, and its ensuing drastic changes in the DOS of the system.

\section{Thermal transitions} 
\label{sec:finitet}
\begin{figure}[t]
\centering
\includegraphics[scale=0.6]{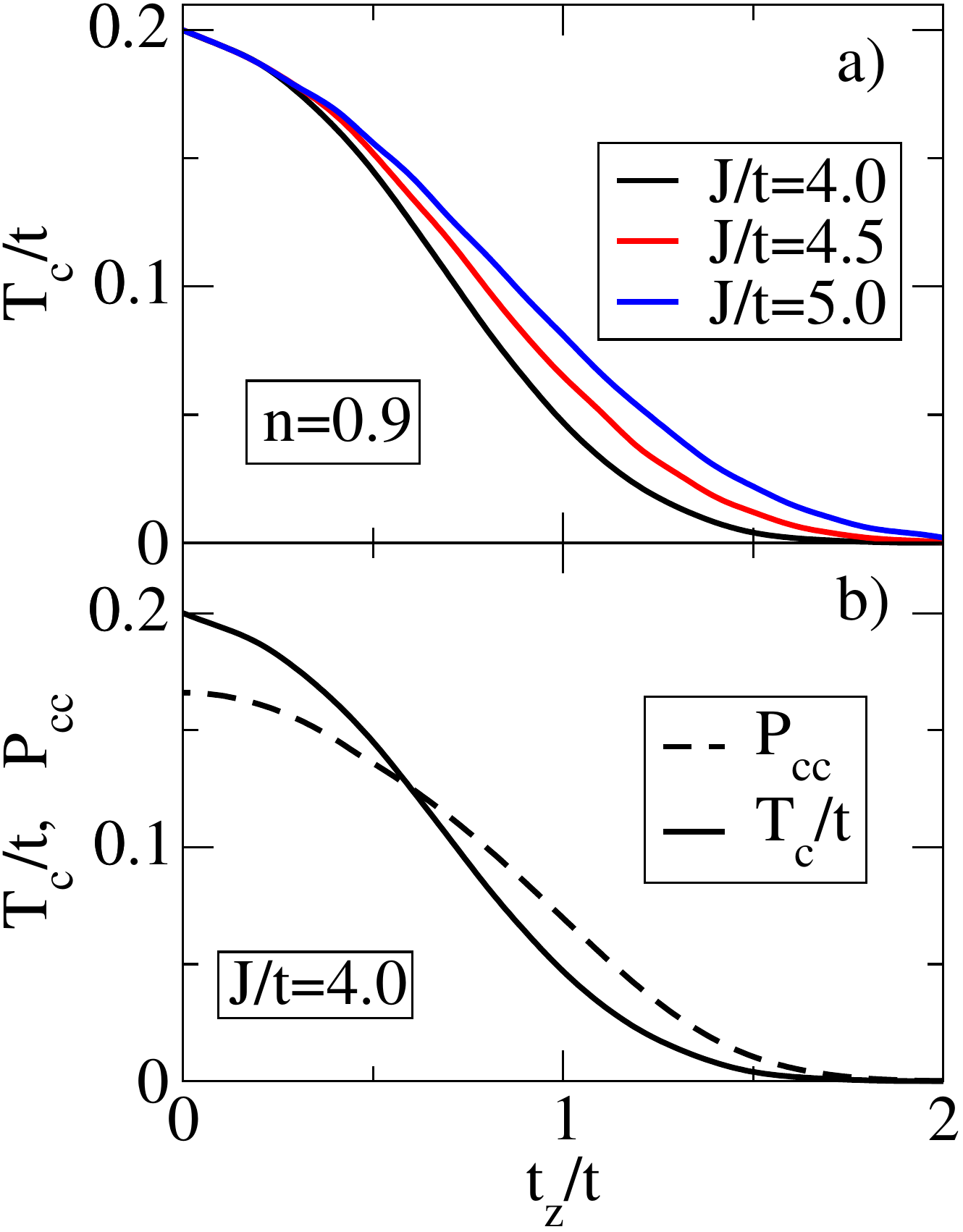} 
\caption{(Color on line) 
Superconducting critical temperature, $T_c/t$, and ground state pairing amplitude, $P_{cc}$, as functions of the interlayer hopping $t_z/t$, for $U/t=2.0$ and $n=0.9$.
In (a), only $T_c$ is shown, for three different values of $J/t$, while in (b) $T_c$ is compared with the ground state pairing amplitude, $P_{cc}$, for a single $J/t$.
}
\label{tcxtz}
\end{figure}
Finally, we briefly discuss thermal effects on the pairing properties by searching for solutions of Eq.\,\eqref{eq:HFthm} at finite temperatures. 
The dominant character of $P_{cc}$ over all other amplitudes was found to be preserved at $T \neq ~0$.
Nonetheless, notwithstanding their differences in magnitude, when plotted as functions of temperature (keeping all other control parameters fixed) all $P_{\alpha\beta}$ were found to vanish at a common critical value, $T_c$.

This fact has two immediate consequences.
Firstly, the existence of a unique $T_{c}$ for all channels adds credence to our claim that pairing occurs amongst the quasiparticles of the Kondo fluid.
Secondly, we can use the behavior of $P_{cc}$ to extract how the critical temperature changes as other parameters are varied. 
For instance, Fig.\,\ref{tcxtz}(a) shows $T_c$ as a function of $t_z/t$ for different values of $J$, all within the Kondo phase. 
We note that increasing $J$ causes an increase in $T_c$, again consistent with the idea that strengthening $df$ singlets enhances the pairing tendency in the $c$-layer. 
The deleterious role of large $d$-$c$ hybridization to superconductivity is also evident at finite temperatures, by the rapid decrease of $T_c$ with $t_z/t$.

Another finite temperature feature worth discussing is whether some manifestation of the proportionality between the superconducting gap function at $T=0$ and the critical temperature is carried over to the present case.
While for the single-band attractive Hubbard model, the gap function is simply proportional to $tP_{cc}$, in the multi-orbital case there is no \textit{a priori} direct relation between the gap function and the pairing amplitudes. 
In order to test this issue, we extract $P_{cc}(t_z/t)$ at $T=0$ and $T_c(t_z/t)$, from Figs.\,\ref{dxtz} and \ref{tcxtz}(a), respectively, and  plot them together in  Fig.\,\ref{tcxtz}(b).
Although they are clearly not proportional to each other, we can say that $T_c$ tracks $P_{cc}(T=0)$.
We have verified that this tracking holds when $P_{cc}(T=0)$ and $T_c$ are plotted against other variables.
For instance, from the data of Fig.\,\ref{dxnc} we conclude that $T_c$ indeed decreases monotonically as the system is doped from half filling.

\section{Conclusions}
\label{sec:concl}

We have studied some proximity effects arising when a normal or superconducting layer is close to a layer of singlets, the latter being either in the insulating or in the metallic state, as controlled by doping. 
To this end, we let a Kondo-lattice layer hybridize (intensity $t_z$) with an additional conducting $c$-layer, in which we can turn on a pairing interaction, $U$. 
Within a Hartree-Fock approximation, we have found that this setup drastically changes the spectral properties of the single-layer Kondo lattice model.
On the magnetic side, $J<J_c(t_z)$, the RKKY interaction is enhanced as a result of the induced antiferromagnetic order on the $c$-layer.  
On the singlet side, hybridization with the $c$-band changes the gap in the $d$-band into a pseudogap (in the sense that it is only non-zero at the Fermi energy). 
The otherwise gapped total density of states now acquires a peak at the Fermi energy.

With the onset of an attractive interaction on the additional metallic layer, superconductivity is now possible for any non-zero $|U|$, due to the drastic changes in the spectral weight at the Fermi energy.
As a result, we have established that pairs may be formed among $c$-, and $f$-electrons, suggesting the occurrence of a \textit{superconducting Kondo phase}. 
We have also established that while the hybridization between the $c$ and $d$-layers is crucial for superconductivity from the spectral point of view, on the other hand it also tends to decrease $T_c$. 
This means that application of uniaxial pressure to increase $t_z$ should be carefully controlled. 
Nonetheless, by adding another mechanism which induces pairing in the $d$-layer (such as local spins fluctuations, Heisenberg coupling of local moments, or magnetic frustration due to next-nearest-neighbor hopping) one could counteract these effects of $t_z$.



\appendix
\section{Hartree-Fock Approximation}
\label{ap:hfa}

The Hartree-Fock approximation for the attractive potential term in Eq.\,\eqref{eq:dEHM}, leads to
\begin{align}
\nonumber
 \sum_{\mathbf{i}}n^{c}_{\mathbf{i}\uparrow}n^{c}_{\mathbf{i}\downarrow}\nonumber &\approx \sum_{\iv}\bigg[   \frac{ n_{c}}{2} \big( c^{\dagger}_{\iv \uparrow} c_{\iv \uparrow} + c^{\dagger}_{\iv \downarrow} c_{\iv \downarrow} \big) - \frac{n^{2}_{c}}{4}-P_{cc}^2
 \\
& +P_{cc} \big( c^{\dagger}_{\iv \uparrow} c^{\dagger}_{\iv \downarrow} + c_{\iv \downarrow} c_{\iv \uparrow} \big) %
  -2  \langle \mathbf{s}^{c}_{\iv} \rangle\! \cdot\! \mathbf{s}^{c}_{\iv} + \langle \mathbf{s}^{c}_{\iv} \rangle\! \cdot\!\langle \mathbf{s}^{c}_{\iv} \rangle \bigg] ,
\label{hamilHub2}
\end{align}
with 
\begin{equation}\label{eq:deltaU}
	P_{cc} = \langle c^{\dagger}_{\iv \uparrow} c^{\dagger}_{\iv \downarrow} \rangle 
				= \langle c_{\iv \downarrow} c_{\iv \uparrow} \rangle
\end{equation}	
being the dimensionless pairing amplitude; for the single-layer attractive Hubbard model  it reduces to the gap parameter, $\Delta \equiv 2 |U| P_{cc}$,
and $n_{c} = \langle c^{\dagger}_{\iv \uparrow} c_{\iv \uparrow}  + c^{\dagger}_{\iv \downarrow} c_{\iv \downarrow} \rangle $ is the electronic density on the $c$ layer, both taken as homogeneous throughout the sites. 
The operator $ \mathbf{s}^{c}_{i} $ is defined as in Eq.\eqref{fermi1}. 
We have checked for the possibility of any residual polarization in the $c$-layer and found none; we can therefore drop the spin terms.
We also do not consider the Hartree terms, $(c^{\dagger}_{\iv \uparrow} c_{\iv \uparrow} + c^{\dagger}_{\iv \downarrow} c_{\iv \downarrow})n_c/2$, since they give rise to spurious unbalanced electronic densities on the layers, with unphysical consequences. 
For instance, in the presence of the Kondo interaction, the Hartree terms stabilize FM phases even close to half filling, $n = (n_c+n_d)/2 = 1$, which is certainly incorrect. 
The reason for this spurious behavior is an artificially strong suppression of the density in the Kondo layer ($n_d$) due to an increase in  $n_c$. 
Nonetheless, we have verified that even without the Hartree terms there is a physically consistent small density imbalance on the layers due to pair formation and the Kondo coupling.
The mean-field Hubbard term then becomes
\begin{equation}
-U\sum_{\mathbf{i}}n^{c}_{\mathbf{i}\uparrow}n^{c}_{\mathbf{i}\downarrow} \approx -U\sum_{i}\bigg[   P_{cc} \big( c^{\dagger}_{i \uparrow} c^{\dagger}_{i \downarrow} + c_{i \downarrow} c_{i \uparrow} \big) - P_{cc}^2 \bigg]
\label{hubbmf}
\end{equation}

Following the procedure presented in Ref.\,\cite{Costa17a}, the Kondo term in Eq.\,\eqref{eq:dEHM}, can be decoupled as
\begin{equation}
\begin{split} 
\sum_{\mathbf{i}} \mathbf{s}^{d}_\mathbf{i}\cdot \mathbf{S}^{f}_{\mathbf{i}} \approx 
&  \sum_{\iv} \bigg[ \mathbf{s}^{d}_{\iv}\! \cdot\! \langle \mathbf{S}^{f}_{\iv} \rangle + \langle \mathbf{s}^{d}_{\iv} \rangle \!\cdot\! \mathbf{S}^{f}_{\iv} \big) -\langle \mathbf{s}^{d}_{\iv} \rangle\! \cdot\! \langle \mathbf{S}^{f}_{\iv} \rangle \\ 
&- \frac{3}{2} \big( V^{0}_{\iv} \langle V^{0}_{\iv} \rangle + \langle V^{0}_{\iv} \rangle V^{0}_{\iv} -\langle V^{0}_{\iv} \rangle \langle V^{0}_{\iv} \rangle \big)\\
& \dfrac{P_{df}}{4}( f^{\dagger}_{\iv\uparrow}d^{\dagger}_{\iv\downarrow}+f^{\dagger}_{\iv\downarrow}d^{\dagger}_{\iv\uparrow}+h.c.)-\dfrac{P_{df}^2}{2}\bigg]
\end{split}
\label{kondomf}
\end{equation}
with the operator
\begin{eqnarray}\label{singlet_hyb}
V^{0}_{\iv}= {V^{0}_{\iv}}^{\dagger} = \frac{1}{2}\sum_{\alpha, \beta = \pm} c^{\dagger}_{\iv \alpha} \mathbb{1}_{\alpha, \beta} f^{\phantom{\dagger}}_{\iv \beta},
\end{eqnarray}
being the singlet hybridization term; we have also investigated a possible influence of triplet hybridization terms \cite{Costa17a}, but they turned out to be irrelevant.

The mean-field values $\langle \mathbf{S}^{f}_{\iv} \rangle$ and $\langle \mathbf{s}^{d}_{\iv} \rangle$ are taken as \cite{Costa17a},
\begin{align}\label{Si}
\langle \mathbf{S}^f_{\iv} \rangle = m_{f} \big[ \cos \left(\mathbf{Q}\!\cdot\! \mathbf{R}_{\iv}\right), \sin \left(\mathbf{Q}\!\cdot\! \mathbf{R}_{\iv}\right), 0 \big]
\end{align}
and
\begin{align}\label{Sci}
\langle \mathbf{s}^{d}_{\iv} \rangle = -m_{d} \big[ \cos \left(\mathbf{Q}\!\cdot\! \mathbf{R}_{\iv}\right), \sin \left(\mathbf{Q}\!\cdot\! \mathbf{R}_{\iv}\right),0 \big],
\end{align}
with 
\begin{align} 
\mathbf{Q}=(q_{x}, q_{y})
\label{Q}
\end{align}
being the magnetic wavevector, and $\mathbf{R}_{\iv}$ the position vector of site $\iv$ on the lattice.

By the same token, the mean values of the hybridization operators are chosen as
\begin{equation}\label{singlet_hyb_mean_v}
\langle V^{0}_{\iv} \rangle = \langle {V^{0}_{\iv}}^{\dagger} \rangle = -V
\end{equation}

Then, our mean-field Hamiltonian, Eq.\,\eqref{eq:HHF}, is obtained by substituting Eqs.\,\eqref{hubbmf} and \eqref{kondomf} in Eq.\,\eqref{eq:dEHM}, using Eqs.\,\eqref{Si}, \eqref{Sci} and \eqref{singlet_hyb_mean_v} for the mean values of the spin and hybridization operators, and performing a discrete Fourier transform, with periodic boundary conditions.
Such Hamiltonian is two-fold degenerate when written in a Nambu spinor basis leading to
\begin{align}
\mathcal{H}_{MF} = \frac{1}{2} \Psi^{\dagger}_{\mathbf{k}} \hat{H}_{\textbf{k}} \Psi_{\mathbf{k}} + const.,
\end{align}
with
\begin{widetext}
\begin{small}
\begin{equation}
\hat{H}_{\textbf{k}}=
\left( \begin{array}{cccccccccccc}
\epsilon_{\textbf{k}} - \mu & -UP_{cc}& 0 & 0 & -t_{z} & 0 & 0 & 0 & 0 & 0 & 0 & 0\\
-UP_{cc} & -\epsilon_{\textbf{-k}} + \mu & 0 & 0 & 0 & t_{z} & 0 & 0 & 0 & 0 & 0 & 0\\
0 & 0 & \epsilon_{\textbf{k+Q}} - \mu  & UP_{cc} & 0 & 0 & -t_{z} & 0 & 0 & 0 & 0 & 0\\
0 & 0 & UP_{cc} & -\epsilon_{\textbf{-k-Q}} + \mu & 0 & 0 & 0 & t_{z} & 0 & 0 & 0 & 0\\
-t_{z} & 0 & 0 & 0 & \epsilon_{k} - \mu & 0 & \frac{1}{2}Jm_{f} & 0 & \frac{3}{4}JV & -\dfrac{J}{4} P_{df} & 0 & 0\\
0 & t_{z}  & 0 & 0 & 0 & -\epsilon_{\textbf{-k}} & 0 & -\frac{1}{2}Jm_{f} & \dfrac{J}{4} P_{df} & -\frac{3}{4}JV & 0 & 0 \\
0 & 0 & -t_{z} & 0 & \frac{1}{2}Jm_{f}  & 0 & \epsilon_{\textbf{k+Q}}& 0  & 0 & 0 & \frac{3}{4}JV & -\dfrac{J}{4} P_{df}\\
0 &0 & 0 & t_{z} & 0 & -\frac{1}{2}Jm_{f}  & 0 & -\epsilon_{\textbf{-k-Q}} 
& 0  & 0 & \dfrac{J}{4} P_{df} & -\frac{3}{4}JV\\
0 & 0 & 0 & 0 & \frac{3}{4}JV & \dfrac{J}{4} P_{df} & 0 & 0 & \epsilon_{f} & 0 & -\frac{1}{2}Jm_{d} & 0\\
0 & 0 & 0 & 0 & -\dfrac{J}{4} P_{df} & -\frac{3}{4}JV & 0 & 0 & 0 & -\epsilon_{f} & 0 & \frac{1}{2}Jm_{d} \\
0 & 0 & 0 & 0 & 0 & 0 & \frac{3}{4}JV  & \dfrac{J}{4} P_{df} & -\frac{1}{2}Jm_{d}  & 0 & \epsilon_{f} & 0\\
0 & 0 & 0 & 0 & 0 & 0 &- \dfrac{J}{4} P_{df} & -\frac{3}{4}JV & 0 & \frac{1}{2}Jm_{d}  & 0 & -\epsilon_{f}
\end{array}\right)
\label{matriz}
\end{equation}
\end{small}
where 
\begin{align}
\Psi^{\dagger}_{k}=(c_{\textbf{k}\uparrow}^{\dagger} c_{-\textbf{k}\downarrow} c_{\textbf{k+Q}\downarrow}^{\dagger}  c_{\textbf{-k-Q}\uparrow} d_{\textbf{k}\uparrow}^{\dagger}  d_{-\textbf{k}\downarrow} d_{\textbf{k+Q}\downarrow}^{\dagger}  d_{\textbf{-k-Q}\uparrow} f_{\textbf{k}\uparrow}^{\dagger}  f_{\textbf{-k}\downarrow}f_{\textbf{k+Q}\downarrow}^{\dagger}  f_{\textbf{-k-Q}\uparrow}),
\end{align}
and
\begin{align}
const.= & N  UP_{cc}^{2} + JN\Big(m_{f}m_{d} +\dfrac{3}{2}V^{2} - \frac{P_{df}^2}{2} \Big) + N[2\mu (n-1) - \epsilon_{f}(n_{f}-1)] 
\end{align}
\end{widetext}
Here, $\mu$ and $\epsilon_{f}$ are included as Lagrange multipliers in order to fix the electronic densities $n$ and $n_f$ respectively. 
The eigenvalues of the Hamiltonian, Eq.\,\eqref{matriz}, are used to obtain the Helmholtz free energy, Eq.\,\eqref{hfe}, and, consequently, the set of nonlinear equations, Eq.\,\eqref{eq:HFthm}, whose solutions yield the sought order parameters.
\subsection{Effective Hamiltonian Diagonalization}

It is instructive to consider the diagonalization of the effective Hamiltonian \eqref{eq:HHF}, at the Fermi level, $\mathbf{k} = \mathbf{k}_F $, in the Kondo phase. The corresponding matrix in the spinor basis $ \Psi^\dagger_\sigma =  (c^\dagger_\sigma \quad d^\dagger_\sigma \quad f^\dagger_\sigma )$ for $U=\mu=\epsilon_f=\epsilon_{\mathbf{k}_F}=0$ is
\begin{equation}
\begin{pmatrix}
0 & -t_z & 0 \\
-t_z & 0 & \dfrac{3}{4}JV \\
0 & \dfrac{3}{4}JV & 0
\end{pmatrix}
\label{efm}
\end{equation}
where the eigenvector related to the Fermi level band, $E_{\mathbf{k}_F} = 0$ is, 
\begin{equation}
|\psi_{FL}\rangle=
\dfrac{1}{\sqrt{9(JV)^2 + 16t_z^2 }}
\begin{pmatrix}
3JV \\
0 \\
4t_z
\end{pmatrix}.
\end{equation}

The spectral weight for the $c$, $d$, and $f$ orbitals are then
\begin{align}
A^c_\sigma({\mathbf{k}_F},0) &=  \quad \dfrac{9J^2V^2}{9J^2V^2 + 16t_z^2 }\\
 A^d_\sigma({\mathbf{k}_F},0) &=  \quad 0\\
 A^f_\sigma({\mathbf{k}_F},0) &=  \quad \dfrac{16t_z^2}{9J^2V^2 + 16t_z^2 }.
\end{align}

\onecolumngrid
\vspace{1cm}
\twocolumngrid

\bibliography{kahmb.bib}

\end{document}